# Reconciling solar and stellar magnetic cycles with nonlinear dynamo simulations

A. Strugarek[1,2]*, P. Beaudoin[1], P. Charbonneau[1], A.S. Brun[2] & J.-D. do Nascimento Jr.[3,4]

[1]*Département de physique, Université de Montréal, C.P. 6128 Succ. Centre-Ville, Montréal, QC H3C-3J7, Canada*
[2]*Laboratoire Astrophysique, Instrumentation, Modélisation (AIM) Paris-Saclay, CEA/Irfu Université Paris-Diderot CNRS/INSU, F-91191 Gif-sur-Yvette*
[3]*Univ. Federal do Rio Grande do Norte (UFRN), Dep. de Física (DFTE), CP 1641, 59072-970, Natal, RN, Brazil*
[4]*Harvard-Smithsonian Center for Astrophysics, Cambridge, Massachusetts 02138, USA*

*Correspondence to: antoine.strugarek@cea.fr

**Abstract**: The magnetic fields of solar-type stars are observed to cycle over decadal periods – 11 years in the case of the Sun. The fields originate in the turbulent convective layers of stars and have a complex dependency upon stellar rotation rate. We have performed a set of turbulent global simulations that exhibit magnetic cycles varying systematically with stellar rotation and luminosity. We find that the magnetic cycle period is inversely proportional to the Rossby number, which quantifies the influence of rotation on turbulent convection. The trend relies on a fundamentally non-linear dynamo process and is compatible with the Sun's cycle and those of other solar-type stars.

The characterization of stellar activity and its dynamo origin has broad applications from exoplanet searches to space weather forecasting. Observational data now allow the determination of absolute luminosities via accurate parallax measurements, rotation through Doppler line broadening and precision photometry, stellar differential rotation through photometry and asteroseismic sounding, and the large-scale spatial structure of stellar photospheric magnetic fields through Zeeman-Doppler imaging. These data complement stellar activity measurements available from long term monitoring programs(*1*, *2*), that showed complex variations of stellar cycle amplitudes and periods as a function of fundamental stellar parameters such as mass, luminosity, rotation and age. The physical understanding of stellar activity is henceforth more complex than suggested by prior interpretation of stellar cycle data through mean-field dynamo theory(*3*-*5*).

Modern global magnetohydrodynamic (MHD) simulations of solar convection and large-scale flows have succeeded in producing in a self-consistent manner large-scale magnetic fields(*6*, *7*), in some cases generating regular, solar-like cyclic magnetic polarity reversals(*8*-*11*). Thus, they are today used to help our physical interpretation of stellar magnetic cycle observations.

We have performed a set of global MHD simulations with the EULAG-MHD code(*12*), using a fixed background stellar structure but covering rotation periods ($P_{rot}$) between 14 and 29 days, and a convective luminosity between 0.2 and 0.6 solar luminosity (see Table S1). The simulated domain consists of a global (i.e. spherical) stellar convection zone with a solar-like aspect ratio (the radius at the bottom of the spherical shell is 70% of the radius at the top $R_{top}$), covering 3.22 density scale-heights with no underlying stable radiation zone. All simulations in the set generate some solar-like features, including: (i) an accumulation of a kilo-Gauss, large-scale axisymmetric magnetic field at the bottom of the convection zone, (ii) regular polarity reversals on a decadal time-scale, reasonably synchronized across hemispheres, (iii) an equatorial propagation of the large-scale magnetic field (see Figure 1), and (iv) solar-like differential rotation (fast equator, slow poles). Some non-solar features were also produced, including the concentration of toroidal magnetic field at mid- rather than low-latitudes, and an irregularly alternating pattern of symmetric and anti-symmetric equatorial parity. This is apparent in Figure 1D, where periods of symmetrical and anti-symmetrical states follow one another. Such parity drifts are understood to reflect the interactions between the two families of dynamo symmetry(*13-16*), which couple in non-linear regime such as the one achieved here.

The magnetic cycle trends in our set of simulations are displayed in Figure 2 (blue circles with error-bars), where two main trends are identified. First, the magnetic cycle period ($P_{cyc}$) is found to decrease in proportion to the rotation rate when the convective luminosity is held constant (Fig 2A). Second, the cycle period also decreases with increasing convective luminosity ($L_{bc}$) at constant rotation rate ($P_{cyc} \propto L_{bc}^{-0.8}$, see Fig 2B, and the supplementary material for their detailed definition). When the results are converted to a non-dimensional form (Fig 2C), or when this luminosity dependency is compensated (Fig 2D), our simulation results follow a single trend that matches the solar cycle.

We compare our results to the growing sample of observed magnetic cycles of distant stars in Fig 2. A first sample was observed with Mt Wilson spectrophotometers(*4*, *5*, *17*). We add to this sample one star that was observed at the Lowell Observatory (*18*), and two stars observed using the High Accuracy Radial Velocity Planet Searcher (HARPS) spectrograph (*19*, *20*). As the luminosities of these stars were not reported in the literature, we calculated it using parallaxes from the Gaia catalogue(*21*, *22*), V magnitudes, and a standard bolometric correction(*23*). The uncertainties on the Gaia parallaxes translate into luminosity uncertainties of less than 10% for most of the stars in the two samples. The samples are composed of stars with very different spectral types (from F to K), and consequently very different convection zone aspect ratios and luminosities. Some stars also exhibit two different cycle periods, in which cases both periods are plotted in Figure 2 and linked by a dashed line.

Historically, two distinct branches in the rotation-cycle period diagram(*3*, *5*) have been favored in the literature. The Sun lies in between these branches (see Fig 2A), requiring conjecture that it is in a transition state between the branches(*24*). A third branch showing an anti-correlation between cycle period and rotation period was also identified for slower rotators(*4*). However, recent observations of solar-type stars seem to indicate a less clear picture that may not reduce to well-defined branches (e.g. see orange diamonds in Fig 2 and also(*25*)). Our simulation results point to only one generic trend, in which the cycle period is inversely proportional to the rotation period. The dependence of the cycle period on the convective luminosity was not considered in earlier analyses of these stellar data, and is shown to be responsible for part of the spread in the rotation-cycle period diagram in Fig 2C and 2D, where the corrected cycle periods of the observed stars then form a broad band inversely proportional to the stellar rotation period, as suggested by our numerical simulations. We highlight three stars (HD 146233, HD 190406 and HD 7615, see Table S2) that are likely to possess a convection zone of depth similar to the sun's. The observational sample still shows a spread around this trend, which is likely due to (i) the varying aspect ratio of the convection zone of the stars in the samples and (ii) the existence of multiple cycle periods for several stars.

The observed cycle period variations with stellar parameters have usually been interpreted through kinematic dynamo models formulated with mean-field theory(*3*, *4*). The two key ingredients in such models are differential rotation and cyclonic turbulence, both resulting ultimately from the action of the Coriolis force on thermally driven convection. In this context, the governing parameter is the Rossby number ($R_o$), which measures the influence of rotation on the system (small Rossby number corresponds to a fast rotating state). The cycle period in our set of simulations is shown to scale as $R_o^{-1}$ (Figure 3), in contrast to dimensional inferences from kinematic, linear mean-field dynamo(*26*), which instead predicts cycle periods proportional to $R_o$.

Our numerical simulations operate in a non-linear regime in which the magnetic force alters the force balance sustaining the large-scale flows(*10*). In Fig 3B we show the systematic acceleration of the differential rotation that modifies the electromotive force to trigger the polarity inversion of the mean azimuthal magnetic field. The amplitude of these fluctuations in the differential rotation is small (~1%), similar to the ones observed on the Sun. A detailed analysis of our simulations (see Figure S8 in the supplementary material) reveals that the torque applied by the large-scale magnetic field controls these modulations. The magnetic cycle period decreases when the amplitude of the differential rotation modulation increases, indicating that non-linear feedback of the Lorentz force on the large-scale differential rotation is driving polarity reversals and setting the cycle period.

Although restricted in the stellar parameter range they span, our simulation results suggest a single trend of cycle period with rotational influence – quantified by the Rossby number – which can accommodate both the Sun and existing stellar data within a single dynamo branch, rather than multiple branches. The scatter about the mean relationship observed between cycle period and rotation rate (Fig 2A) can be partly attributed to the sensitive dependence of the cycle period on luminosity. The remaining scatter remains to be explained and could originate from structural factor such as the exact depth of the convection zone or the exact shape of the differential rotation, which have not been explored yet. These considerations reinstate the Sun to the status of an ordinary solar-type star, and a robust calibration point for stellar astrophysics.

**References**

1. S. L. Baliunas et al., Astrophys. J. 438, 269–287 (1995).
2. J. C. Hall, G. W. Lockwood, B. A. Skiff, Astronom. J. 133, 862–881 (2007).
3. R. W. Noyes, N. O. Weiss, A. H. Vaughan, Astrophys. J. 287, 769–773 (1984).
4. S. H. Saar, A. Brandenburg, Astrophys. J. 524, 295–310 (1999).
5. E. Bohm Vitense, Astrophys. J. 657, 486–493 (2007).
6. A. S. Brun, R. A. Garcia, G. Houdek, D. Nandy, M. Pinsonneault, Space Sci. Rev. 196, 303–356 (2015).
7. H. Hotta, M. Rempel, T. Yokoyama, Science 351, 1427–1430 (2016).
8. M. Ghizaru, P. Charbonneau, P. K. Smolarkiewicz, Astrophys. J. 715, L133–L137 (2010).
9. P. J. Käpylä, M. J. Mantere, A. Brandenburg, Astrophys. J. 755, L22 (2012).
10. K. Augustson, A. S. Brun, M. Miesch, J. Toomre, Astrophys. J. 809, 149 (2015).
11. L. D. V. Duarte, J. Wicht, M. K. Browning, T. Gastine, Mon. Rot. R. Astron. Soc. 456, 1708–1722 (2016).
12. P. K. Smolarkiewicz, P. Charbonneau, J. Comput. Phys. 236, 608–623 (2013).
13. A. Brandenburg, F. Krause, R. Meinel, D. Moss, I. Tuominen, Astronom. Astrophys. 213, 411–422 (1989).
14. D. Gubbins, K. Zhang, Phys. Earth Planet. Inter. 75, 225–241 (1993).
15. S. M. Tobias, Astronom. Astrophys. 322, 1007–1017 (1997).
16. M. L. DeRosa, A. S. Brun, J. T. Hoeksema, Astrophys. J, 757, 96 (2012).
17. R. W. Noyes, L. W. Hartmann, S. L. Baliunas, D. K. Duncan, A. H. Vaughan, Astrophys. J. 279, 763–777 (1984).
18. J. C. Hall, G. W. Henry, G. W. Lockwood, Astronom. J. 133, 2206–2208 (2007).
19. M. Mayor et al., The Messenger (ISSN0722-6691). 114, 20–24 (2003).
20. R. F. Díaz et al., Astronom. Astrophys. 585, A134 (2016).

21. Collaboration Gaia et al., Astronom. Astrophys. 595, A1 (2016).

22. Collaboration Gaia et al., Astronom. Astrophys. 595, A2 (2016).

23. G. Torres, Astronom. J. 140, 1158–1162 (2010).

24. T. S. Metcalfe, R. Egeland, J. van Saders, Astrophys. J. 826, L2 (2016).

25. V. See et al., Mon. Rot. R. Astron. Soc. 462, 4442–4450 (2016).

26. R. W. Noyes, N. O. Weiss, A. H. Vaughan, Astrophys. J. 287, 769–773 (1984).

27. F. Ochsenbein, P. Bauer, J. Marcout, Astronom. Astrophys. Suppl. Ser. 143, 23–32 (2000).

28. Astroquery package, (available at http://astroquery.readthedocs.io/en/latest/#).

29. K. H. Schatten, J. M. Wilcox, N. F. Ness, Sol. Phys. 6, 442–455 (1969).

30. S. T. Fletcher et al., Astrophys. J. 718, L19–L22 (2010).

31. R. Simoniello et al., Astrophys. J. 765, 100 (2013).

32. R. K. Ulrich, T. Tran, Astrophys. J. 768, 189 (2013).

33. S. R. Lantz, Y. Fan, Astrophys. J. Suppl. Ser. 121, 247–264 (1999).

34. J. A. Domaradzki, Z. Xiao, P. K. Smolarkiewicz, Phys. Fluids 15, 3890–3893 (2003).

35. A. Strugarek et al., Adv. Space Res. 58, 1538–1553 (2016).

36. C. A. Jones et al., Icarus. 216, 120–135 (2011).

37. B. P. Brown, M. K. Browning, A. S. Brun, M. S. Miesch, J. Toomre, Astrophys. J. 711, 424–438 (2010).

38. A. S. Brun et al., Astrophys. J. 836, 192 (2017).

39. N. A. Featherstone, M. S. Miesch, Astrophys. J. 804, 67 (2015).

40. P. Charbonneau, Living Review in Solar Physics 7, 3 (2010).

41. L. Jouve, B. P. Brown, A. S. Brun, Astronom. Astrophys. 509, 32 (2010).

42. F. van Leeuwen, Astrophys. Space Sci. Lib. 350 (2007), doi:10.1007/978-1-4020-6342-8.

43. S. G. Sousa et al., Astronom. Astrophys. 487, 373–381 (2008).

44. S. Boro Saikia et al., Astronom. Astrophys. 594, A29 (2016).

45. N. R. Lomb, Astrophys. Space Sci. 39, 447–462 (1976).

46. J. D. Scargle, Astrophys. J. 263, 835–853 (1982).

**Acknowledgments:** We thank P. Smolarkiewicz for help and advice on using the EULAG code and J.F. Cossette for discussions about the modeling of convection inside stars. We also thank R. Garcia for many discussions on stellar rotation. The authors acknowledge support from Canada's Natural Sciences and Engineering Research Council. This work was also partially supported by the INSU/PNST, the ANR 2011 Blanc Toupies, the ERC Grant STARS2 207430, CNES Solar Orbiter and PLATO grants.

This work made use of the VizieR database(*27*), through the Astroquery package(*28*).

This work has made use of data from the European Space Agency (ESA) mission Gaia (https://www.cosmos.esa.int/gaia), processed by the Gaia Data Processing and Analysis Consortium (DPAC, https://www.cosmos.esa.int/web/gaia/dpac/consortium).

The EULAG-MHD code can be accessed at http://www.astro.umontreal.ca/~paulchar/grps/eulag-mhd.html (subject to US export restrictions). The output of the simulations presented in this work are downloadable at http://www.astro.umontreal.ca/sun/.

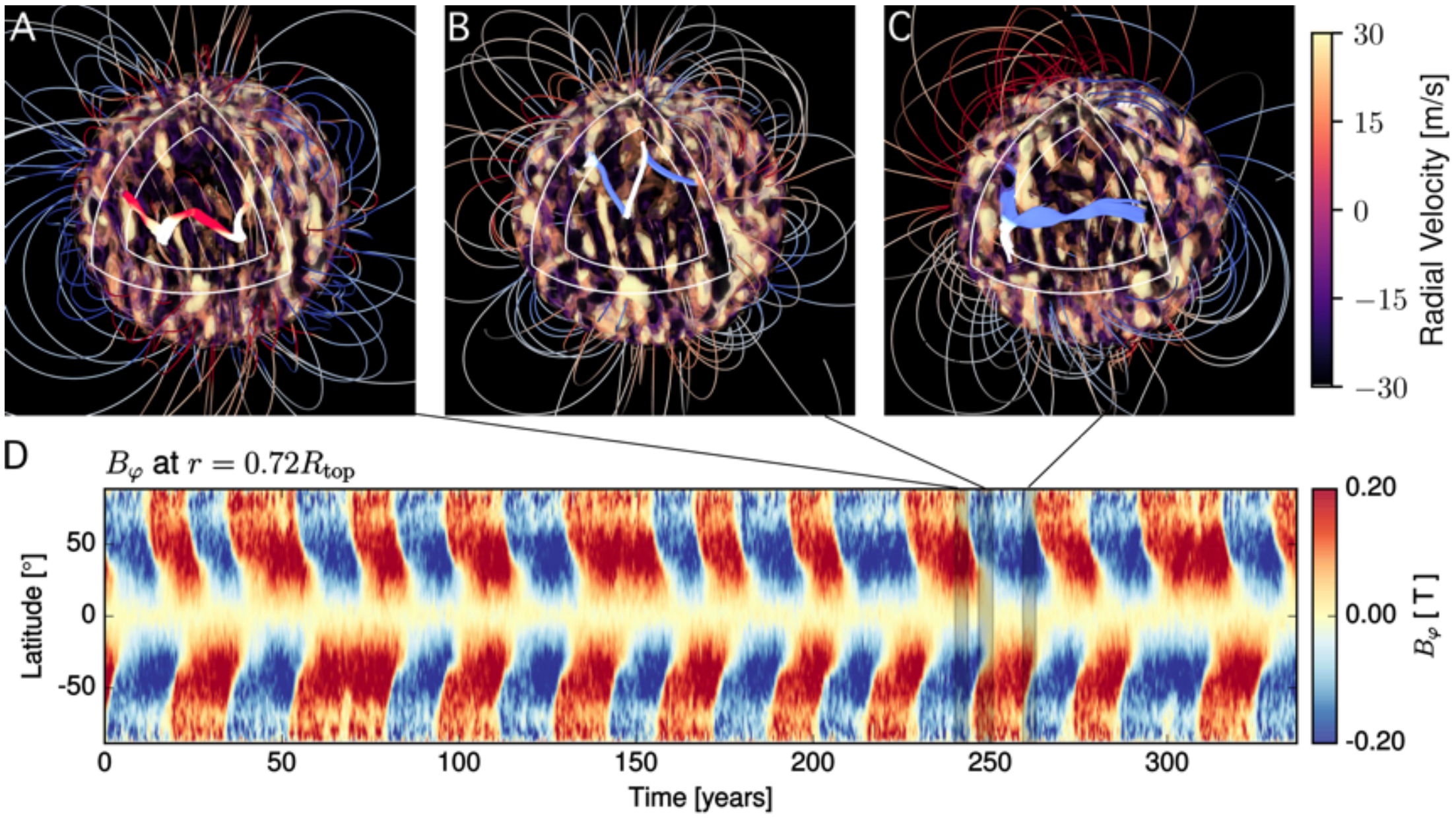

**Fig. 1 A non-linear, global magnetic cycle.** (A-C) Snapshots of a representative three-dimensional non-linear simulation of a regular magnetic cycle. White (positive) and purple

(negative) volumes represent the radial velocity (in m/s) of the convective flow. A half sector of the spherical shell has been cut out to display the large scale magnetic field lines (averaged over 50 rotation periods) buried in the convection zone (the red/blue coloring of the magnetic tubes labels positive/negative azimuthal magnetic field). The magnetic field lines extending beyond the simulation domain are derived using a standard potential field extrapolation(*29*). (D) Longitudinal average of the azimuthal component of the magnetic field ($B_\varphi$) as a function of latitude and time at depth r=0.72 $R_{top}$.

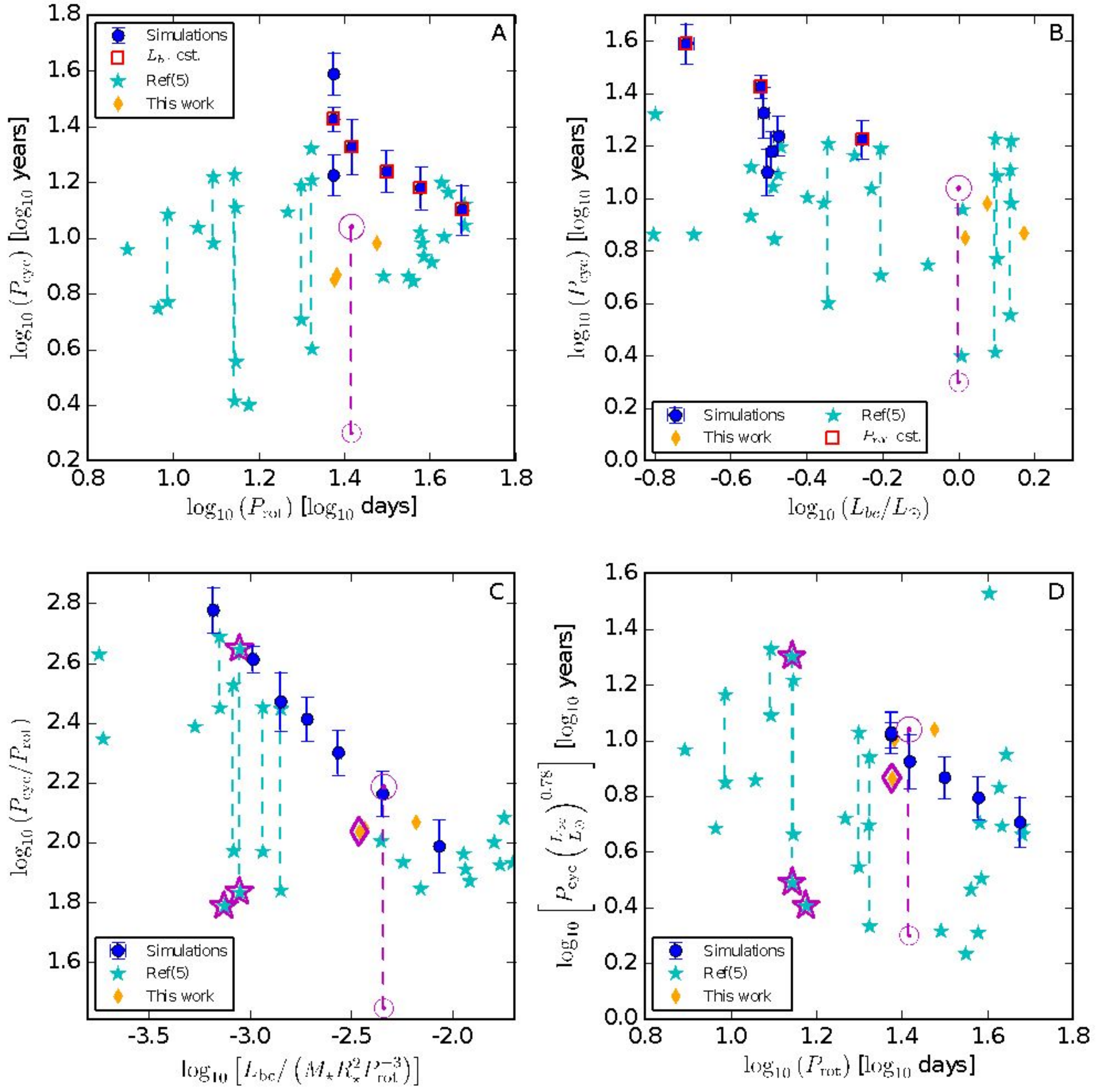

**Fig. 2 Trends of the magnetic cycle period**. The cycle half-period (11 and 2 years for the Sun(*30-32*), magenta ⊙ symbol) are plotted against rotation period (A) and stellar luminosity (B) for our set of simulations (blue circles) and two observed samples of stars (cyan stars and orange diamonds)(*4*, *5*). Panel (C) represents the same quantities in a normalized way, with the cycle period normalized to the rotation period and the luminosity normalized to $M_\star R_\star^2 P_{rot}^{-3}$ ($M_\star$ and $R_\star$ are the mass and the radius of the stars). The dependence of the cycle period upon the stellar convective luminosity observed in our set of simulations (panel B) is factored out in panel D. In all panels, the best fit (using orthogonal distance regression) of our simulation data is shown by the grey dashed lines.

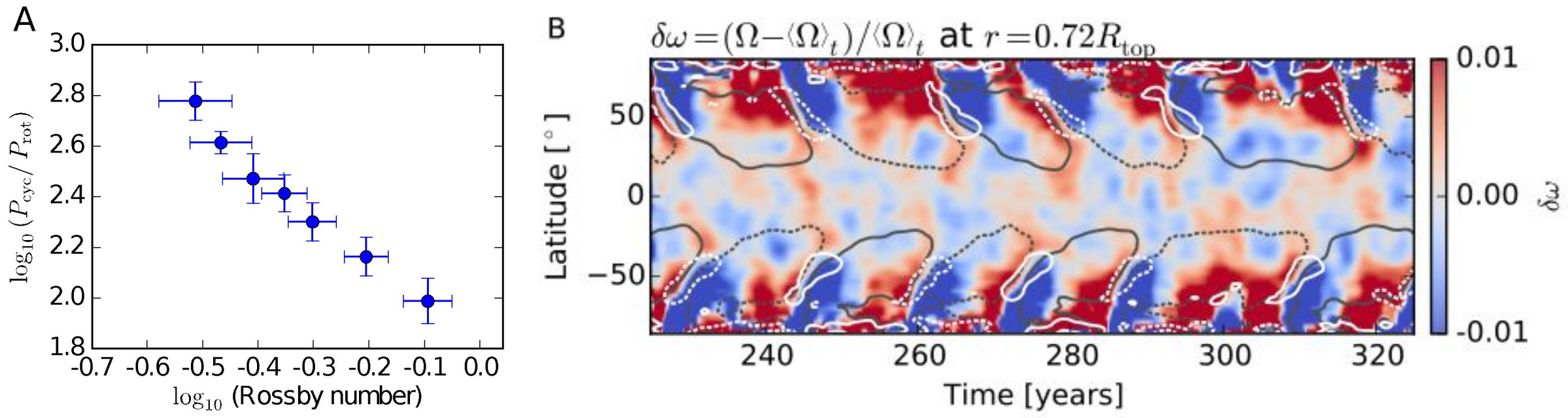

**Fig. 3 Interpretation of the non-linear convective dynamo**. (A) Magnetic cycle period (normalized to the rotation period) as a function of the local Rossby number. The scaling law indicated by the dashed grey line is fitted with orthogonal distance regression. (B) Relative variation in time (δω) of the differential rotation as a function of latitude and time at depth r=0.72 $R_{top}$ (Ω is the differential rotation defined as the azimuthal velocity divided by the cylindrical radius, and $<>_t$ stands for the temporal average). The iso-contours of the mean azimuthal magnetic field at ±0.1 Tesla are shown in grey. The contribution of shearing by differential rotation to the mean electro-motive force is shown as white contours. All fields in panel B have been smoothed with a running average of 4 years using a hann window.

**Supplementary Materials:**

Materials and Methods

Figures S1-S9

Tables S1 and S2

References (*33-46*)

The EULAG code

The Eulerian-Lagrangian (EULAG-MHD) code solves the anelastic form of the MHD equations(*33*) without explicit dissipative terms. Numerical stability is enforced through the dissipation provided by the underlying advection algorithm(*12*), which effectively acts as an implicit subgrid model(*34*, *35*). In our simulations, the domain consists of a convective layer with no underlying stable zone. The models are set up exactly as in (*35*, *36*) for the hydrodynamic part. We reach a density scale height at the top of the domain which corresponds to 96% of the solar radius. The parameters that changed compared to (*35*) are listed in table S1. The bottom magnetic boundary condition has a drastic impact on the dynamo state the system reaches(*37*). As a result, we choose for the bottom spherical boundary a perfectly conducting stress-free wall, which acts as a deep strongly stratified conducting layer.
The top of the simulation domain is also a stress-free wall, and the magnetic field is constrained to be radial in order to mimic the connection to a stellar chromospheric layer. Convective motions are driven through a volumetric heating/cooling driving the thermal structure towards a mildly superadiabatic ambient stratification(*12*). Under such a thermal forcing setup, the transport of heat (the convective luminosity) mediated by convective motions is an output of the simulation and is computed a posteriori. We choose here models carrying only a fraction of the solar luminosity to ensure that the Rossby number of our models remain lower than one (see Table S1). This allows our models to have qualitatively similar differential rotation profiles (see Table S1 and (*38*)).

The numerical method behing EULAG-MHD uses implicit dissipation that render the estimation of classical adimensional MHD numbers (Prandtl, Reynolds, Rayleigh numbers) difficult. Using a similar grid, the implicit viscosity ($\nu$) and thermal diffusion ($\kappa$) were estimated to be of $10^{12}$ $cm^2/s$ the order in hydrodynamical simulations(*35*), with a Prandtl number ($\nu/\kappa$) of order 1 away from boundaries.

| **Model** | $\boldsymbol{\Omega_\star}$ $[\boldsymbol{\Omega_\odot}]$ | $\boldsymbol{\Delta S}$ **[erg/g/K]** | $\boldsymbol{L_\star}$ $[\boldsymbol{L_\odot}]$ | $\boldsymbol{P_{cyc}}$ **[years]** | $\boldsymbol{R_o}$ | $\boldsymbol{R_{os}}$ |
|---|---|---|---|---|---|---|
| **1** | 1.1 | $1.0 \times 10^4$ | $0.29 \pm 0.01$ | $26.7 \pm 2.9$ | $0.34 \pm 0.05$ | $0.28 \pm 0.09$ |
| **2** | 1.1 | $0.8 \times 10^4$ | $0.20 \pm 0.01$ | $39.0 \pm 7.5$ | $0.31 \pm 0.05$ | $0.20 \pm 0.06$ |
| **3** | 1.1 | $1.5 \times 10^4$ | $0.61 \pm 0.02$ | $16.9 \pm 3.1$ | $0.44 \pm 0.04$ | $0.37 \pm 0.13$ |
| **4** | 0.83 | $1.0 \times 10^4$ | $0.32 \pm 0.01$ | $17.3 \pm 3.3$ | $0.50 \pm 0.05$ | $0.44 \pm 0.15$ |
| **5** | 0.55 | $1.0 \times 10^4$ | $0.32 \pm 0.01$ | $12.6 \pm 2.9$ | $0.81 \pm 0.09$ | $0.68 \pm 0.21$ |
| **6** | 0.69 | $1.0 \times 10^4$ | $0.33 \pm 0.01$ | $15.2 \pm 2.9$ | $0.62 \pm 0.06$ | $0.55 \pm 0.17$ |
| **7** | 1 | $1.0 \times 10^4$ | $0.30 \pm 0.01$ | $21.3 \pm 5.3$ | $0.39 \pm 0.05$ | $0.31 \pm 0.10$ |

**Table S1 Simulation parameters and results.** The input parameters of the models (rotation rate $\Omega_\star$, entropy contrast through the convection zone $\Delta S$) are reported for each model. The other quantities computed a posteriori for each model are the convective luminosity $L_{bc}$ (Eq S1), the cycle period $P_{cyc}$ (see text), the fluid Rossby number $R_o$ (Eq S2) and the stellar Rossby number $R_{os}$ (Eq S3).

Set of numerical simulations

We display summary figures of all the numerical simulations in our study in Figures S1-S7. Figure S5 shows the reference simulation shown in Figure 1 (case 5). The differential rotation ($\Omega$, panels A) is solar-like (slow poles, fast equator) in all cases. It strengthens when the rotation rate increases. It weakens at the equator when the luminosity increases, but increases overall due to a strong decrease of $\Omega$ at the poles. The strength of the meridional flow (panels B, represented by the streamfunction $\Psi_{MC}$) is correlated with the amplitude of the differential rotation(*39*). It is composed by one cell at high latitudes, and multiple stacked cells at the equator. The convective flows (panels C, middle of the convection zone) are structured in elongated shapes near the equator, testifying of the influence of rotation on the turbulent flows. The faster the model rotates, the more elongated are the structure. The magnetic cycle appears in panels D and E showing cuts of the azimuthally averaged azimuthal component of the magnetic field. The magnetic field is symmetric (quadrupolar) with respect to the equator in some cases, while in other cases a beating between symmetric (quadrupolar) and anti-symmetric (dipolar) structures occurs. The convective luminosity profile throughout the convection zone is shown in panel F. It varies with radius due to the convection-forcing scheme. We choose to define the convective luminosity of our models by averaging it over the gray band in panel F, which is where the oscillatory dynamo primarily resides. The trends identified in this paper are robust with respect to his particular definition as the luminosity varies by approximately the same amount at all radii from model to model. Finally, the Fourier spectrum of the azimuthal field (panel G) gives a first estimate of the cycle period of each model. We display in Table S1 the main parameters and results of all the models considered in this work.

Estimation of the cycle period

We estimate the magnetic cycle period in our numerical simulations by running a Fourier transform on the longitudinally averaged azimuthal component of the magnetic field at all points in radius and latitude. The main peak of the Fourier transform is stored at each point, along with a period bandwidth defined as a band around the main peak for which the Fourier transform is larger than 10% of its peak value. A probability density function is formed with all the stored peaks, and the maximum of the distribution defines the main cycle period of the simulation. The uncertainty on the cycle period is calculated as the maximal bandwidth associated with the cycle-period. Note that the cycle period shown in Figure 2 and Tables S1 and S2 are actually the half-cycle period (i.e. 11 years for the Sun).

Estimation of the convective luminosity

The convective luminosity is estimated by computing the azimuthal and latitudinal average of the convective heat flux averaged over several cycle periods such as

$$L(r) = 4\pi\, r^2 \rho\, c_p < v_r T' >, \quad \text{(S1)}$$

where $T'$ are the temperature fluctuations, $v_r$ the radial velocity, $\rho$ the density, $c_p$ the specific heat of the plasma at constant pressure, and <> denotes the average over the sphere of radius $r$ and over time (here over several magnetic cycles). The convective luminosity at the base of the convection zone $L_{bc}$ is defined here as the average of the convective luminosity $L(r)$ over the radial interval [0.75 $R_{top}$, 0.8 $R_{top}$], in the region where dynamo action is taking place and sufficiently removed from the lower boundary.

Estimation of the Rossby number

The Rossby number quantifies the relative importance of non-linear advection to the Coriolis acceleration in the Navier-Stokes equations of fluid dynamics expressed in a rotating frame of reference. This fluid Rossby number is defined by

$$R_o = | \nabla \times \boldsymbol{v} | / ( 2\, \Omega_\bigstar ), \qquad \text{(S2)}$$

where $\boldsymbol{v}$ is the plasma velocity and $\Omega_\bigstar$ the rotation rate of the model. We average the Rossby number over time and longitude, over the same radial range [0.75 $R_{top}$, 0.8 $R_{top}$] as the convective luminosity, and over a latitudinal wedge of 140° centered on the equator. The uncertainty in the Rossby number is taken to be its standard deviation over the averaging surface in the radius-latitude plane (see table S1).

In the context of cool stars, another Rossby number (the stellar Rossby number) is often also defined as a ratio of two timescales (see e.g. Appendix B in(*38*)), the rotation period of the star ($P_{rot} = 2\pi/ \Omega_\bigstar$), and the convective turnover time at the base of the convection zone of the star ($\tau_c$). We also computed this stellar Rossby number in the middle of the convection zone as

$$R_{os} = P_{rot} / \tau_c\,, \qquad \text{(S3)}$$

The the convective turnover-time is calculated as the ratio of the depth of the convection zone to the root-mean-square radial velocity at the middle of the convection zone, over the same latitudinal range as $R_o$. Both Rossby numbers ($R_o$ and $R_{os}$, see table S1) are a measure of the rotational influence on the convective dynamics and result in the same generic trends regarding the magnetic cycle period.

Non-linear feedbacks in the cyclic dynamo

We display in Figure S8 the origin of the non-linear feedback loop in the simulated cyclic dynamo for case 5. The non-linear feedback occurs as follows. The large-scale magnetic field (black line in the upper panel, label A) applies a Lorentz force that is positive when the field is maximum (blue line in the lower panel, label B). This Lorentz force locally increases the differential rotation, which in turns inverts its local latitudinal gradient (black line in the lower panel, label C). As a result, the sign of the mean electro-motive force switches (blue line in the upper panel, label D). Note that the latitudinal field (green line in the upper panel) changes sign later on in the cycle (label E), showing unambiguously that the change of sign of the mean electromotive force indeed originates from a change of sign of the latitudinal gradient of differential rotation. The large-scale magnetic field then collapses due to this opposite electromotive force, and grows again with an opposite sign (label F). This non-linear feedback then continues on and on and determines the length of the magnetic cycle.

Mean-field dynamo models and magnetic cycle period trends

Mean-field models of solar and stellar dynamos come in many flavors that are known to lead to very different scaling laws(*40*). Standard α-Ω dynamos predict a cycle period proportional to the rotation rate of the star, while Babcock-Leighton models generally find the opposite trend when the meridional circulation amplitude is parametrized using three-dimensional numerical simulations(*41*). In all mean-field cases, the basic parameters of the models (differential rotation Ω, meridional circulation, ohmic dissipation, turbulent electromotive force) need to be scaled a priori with the basic stellar parameters (rotation rate, luminosity). As a result, different scaling laws can be realized depending on both the detailed dynamo model considered, and the scaling chosen for their basic ingredients. Note that even though the Babcock-Leighton dynamo mechanism is very different from the one presented in this work, it generally predicts a cycle-period trend that is compatible with the one found in our three-dimensional non-linear simulations given the known dependency of the meridional circulation with the stellar rotation rate.

Estimation of the luminosity of the observed stars

We follow a standard procedure(*23*) for estimating the luminosity of a star in the two samples used in this study. The effective temperatures of the stars were originally published along the reported cycle periods(*5*). The luminosity $L_\star$ of a star is approximated through

$$\log_{10} ( L_\star / L_\odot ) = -0.4 ( m_V - 5 \operatorname{Log}_{10}(d/10) + BC_V - M_{bol}^{\odot}) , \quad (S4)$$

where $m_V$ is the apparent magnitude in the *V* band, *d* the distance between the Sun and the star (in parsecs), $M_{bol}^{\odot}$ the bolometric magnitude of the Sun and $BC_V$ the bolometric correction, which is an empirical function of the effective temperature(*23*). The most distant star in the sample is located 30.41 pc away (HD 81809, see Table S2), which allow us to neglect luminosity corrections due to interstellar extinction. The distance *d* and $m_V$ were taken from the latest Gaia catalogue(*21*, *22*). We also compared with values from the Hipparcos catalogue(*42*) but did not find any substantial change in the results. We report the calculated luminosities in Table S2.

| **Name** | $P_{rot}$ **[days]** | $P_{cyc}$ **[years]** | $P_{cyc}(2)$ **[years]** | $T_{eff}$ **[K]** | $L_\star$ $[L_\odot]$ | **Mass** $[M_\odot]$ | **Radius** $[R_\odot]$ | **d [pc]** | **Source** |
|---|---|---|---|---|---|---|---|---|---|
| HD 100180 | 14 | 12.9 | 3.6 | 5951 | 1.37 | 0.94 | 1.11 | 23.33 | Ref(5,21) |
| HD 114710 | 12.4 | 16.6 | 9.6 | 5920 | 1.38 | 0.75 | 1.13 | 9.13 | Ref(5,21) |
| HD 78366 | 9.7 | 12.2 | 5.9 | 5856 | 1.26 | | 1.1 | 19.19 | Ref(5,21) |
| HD 190406 | 13.9 | 16.9 | 2.6 | 5825 | 1.25 | 0.79 | 1.11 | 17.77 | Ref(5,21) |
| HD 1835 | 7.8 | 9.1 | | 5670 | 1.03 | 0.62 | 1.06 | 20.86 | Ref(5,21) |
| HD 20630 | 9.2 | 5.6 | | 5609 | 0.83 | 0.93 | 0.97 | 9.14 | Ref(5,21) |
| HD 76151 | 15 | | 2.52 | 5639 | 1.02 | 1.04 | 1.07 | 17.39 | Ref(5,21) |
| HD 152391 | 11.4 | 10.9 | | 5373 | 0.59 | | 0.9 | 17.25 | Ref(5,21) |
| HD 81809 | 40.2 | 8.2 | | 5258 | 6.1 | 0.93 | 3.0 | 30.41 | Ref(5,21) |
| HD 103095 | 31 | 7.3 | | 5401 | 0.2 | 1.01 | 0.52 | 9.09 | Ref(5,21) |
| HD 115404 | 18.5 | 12.4 | | 4875 | 0.33 | 1.52 | 0.82 | 11.07 | Ref(5,21) |
| HD 149661 | 21.1 | 16.2 | 4 | 5145 | 0.45 | 1.05 | 0.86 | 9.75 | Ref(5,21) |
| HD 156026 | 21 | 21 | | 4335 | 0.16 | | 0.71 | 5.97 | Ref(5,21) |
| HD 165341 | 19.9 | 15.5 | 5.1 | 5091 | 0.62 | 1.04 | 1.03 | 5.08 | Ref(5,21) |
| HD 4628 | 38.5 | 8.6 | | 5036 | 0.28 | 0.72 | 0.71 | 7.45 | Ref(5,21) |
| HD 10476 | 38.2 | 9.6 | | 5145 | 0.44 | 0.78 | 0.84 | 7.53 | Ref(5,21) |
| HD 160346 | 36.4 | 7 | | 4824 | 0.33 | 0.86 | 0.83 | 11.00 | Ref(5,21) |
| HD 201091 | 35.4 | 7.3 | | 4307 | 0.16 | | 0.72 | 3.49 | Ref(5,21) |
| HD 201092 | 37.8 | 10.5 | | 3867 | 0.12 | 0.62 | 0.79 | 3.49 | Ref(5,21) |
| HD 3651 | 44 | 14.6 | | 5118 | 0.53 | 0.74 | 0.94 | 11.06 | Ref(5,21) |
| HD 16160 | 48 | 13.2 | | 4772 | 0.28 | 0.81 | 0.79 | 7.18 | Ref(5,21) |
| HD 26965 | 43 | 10.1 | | 5201 | 0.4 | 0.67 | 0.79 | 4.98 | Ref(5,21) |
| HD 32147 | 48 | 11.1 | | 4529 | 0.33 | 0.62 | 0.94 | 8.71 | Ref(5,21) |
| HD 166620 | 42.4 | 15.8 | | 5063 | 0.34 | | 0.77 | 11.02 | Ref(5,21) |
| HD 10180 | 24.1 | 7.4 | | 5911 | 1.49 | 1.02 | 1.10 | 39.02 | Ref(21) |
| HD 146233 | 23.8 | 7.1 | | 5818 | 1.04 | 0.90 | 1.01 | 13.90 | Ref(18,21) |
| HD 1461 | 29.9 | 9.6 | | 5765 | 1.19 | 1.01 | 1.01 | 23.25 | Ref(20,21) |

**Table S2 Parameters of the stellar samples**. The rotation period, cycle periods and effective temperature were taken from (*5*, *43*), except for the three last stars. The cycle period of HD 146233 is taken from (*18*), and the cycle period of HD 1461 from (*20*). The cycle period of HD 10180 is analyzed in this paper. The luminosity is calculated with Eq S4. The mass is taken from (*43*) (when available), and the radius is deduced from Eq S5. The distance is taken from the latest Gaia catalogue (*21*, *22*).

Estimation of the mass and radius of stars

When available, we used the stellar mass $M_\star$ deduced from observations with the High Accuracy Radial Velocity Planet Searcher (HARPS) spectrograph(*43*), which we reported in Table S2. We computed the stellar radius $R_\star$ using the deduced luminosity and published effective temperature $T_{eff}$(*5*, *43*) using the standard formula

$$\log_{10} ( R_\star / R_\odot ) = 0.5 \log_{10} ( L_\star / L_\odot ) - 2 \log_{10} (T_{eff} / T_\odot) \quad . \qquad \text{(S5)}$$

Magnetic activity cycle of HD 10180

We used the HARPS southern archive data to study the variations in the Ca II H&K emission cores measurements of HD 10180, that span 4084 days. The Ca II H & K line cores bands are centered at 3933.664 Å (K) and 3968.470 Å (H). We calculate the HK index Log $R'_{HK}$ defined in (*18*) and compute a Lomb-Scargle periodogram to estimate the magnetic activity cycle period of HD 10180 (see Figure S9). We find a cycle period of 7.3 ± 1.2 years, which is reported in Table S2.

Magnetic cycles and chromospheric activity cycles

An important distinction must be made regarding the reported activity cycles of distant stars. Most of the published cycles are deduced from fluctuation in the activity tracers of the star (e.g. Ca II $R'_{HK}$). As of today, a regular cycle directly observed on the magnetic field of a star have only been reported for two stars: the Sun and 61 Cyg A (HD 201901) (*44*). Both stars are taken into account in this work. For both of them the activity cycle and the magnetic cycle are very well correlated. We note here nonetheless that the vast majority of the cycles reported in the literature are activity cycles, and still lack a true validation through spectropolarimetry(*25*).

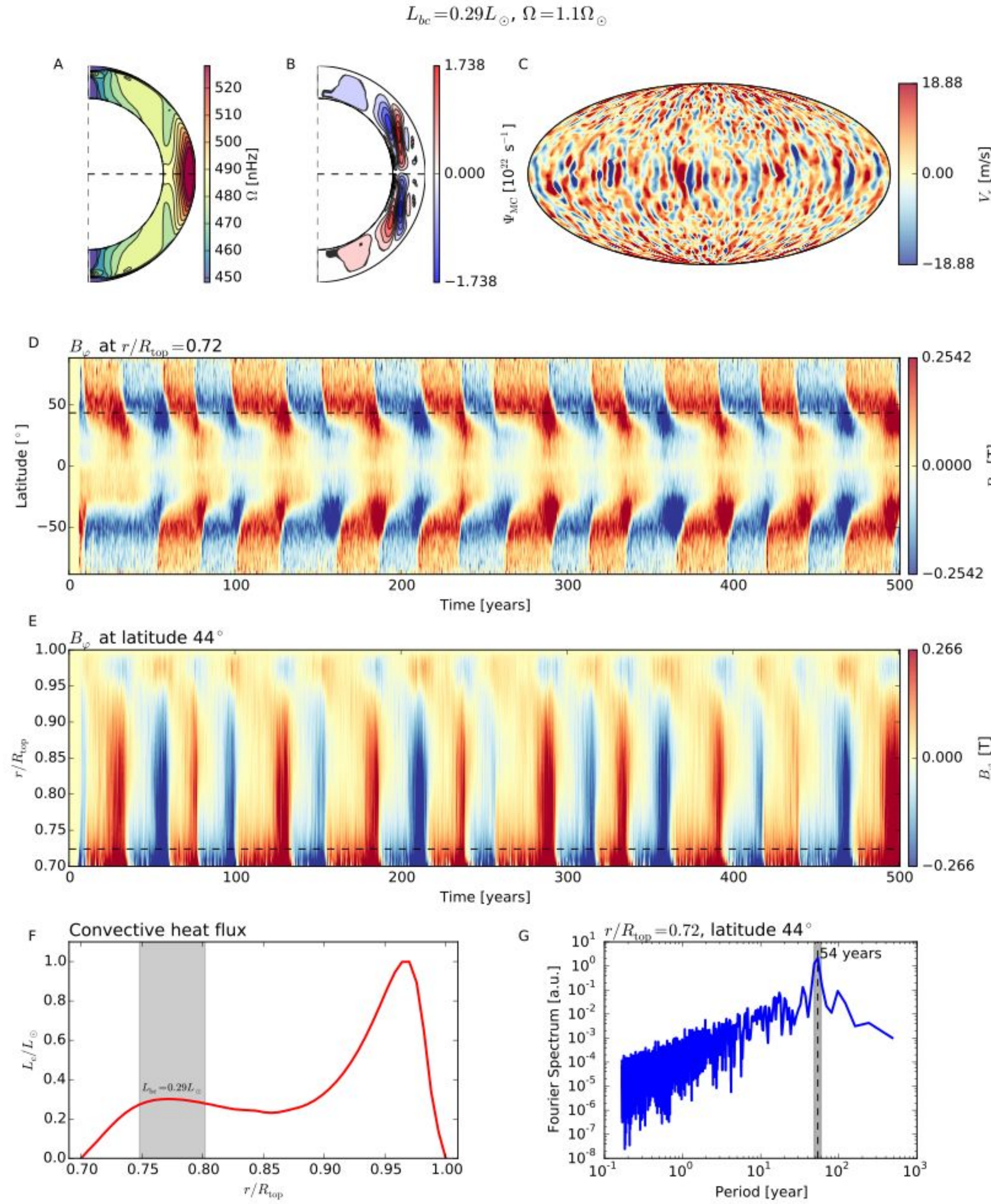

**Fig. S1 Summary of a cyclic numerical simulation (Model 1).** (A) Differential rotation profile ($\Omega = \Omega_\star + v_\varphi/(r \sin\theta)$, with θ the co-latitude) averaged over time and longitude. (B) Meridional circulation streamlines averaged over time and longitude. Clockwise circulations are shown in red, anti-clockwise in blue. (C) Radial velocity on a Mollweide projection on the sphere at depth r=0.75 $R_{top}$. Red denotes upflows and blue downflows. (D) Azimuthal component of the magnetic field averaged over longitude as a function of time and latitude at depth r=0.75 $R_{top}$ (near the base of the convection zone). (E) Azimuthal component of the magnetic field averaged over longitude as a function of time and normalized radius at latitude 44°. (F) Convective luminosity as a function of depth. The grey band corresponds to the depth of large magnetic fields where we choose to estimate the convective luminosity used in Figure 2. (G) Fourier spectrum of the azimuthal component of the magnetic field at depth r=0.75 $R_{top}$ and latitude 44°

(dashed black lines in panels D and E). The peak of the spectrum shows the magnetic cycle period. The magnetic cycle periods shown in Figure 2 and 3 are calculated with Fourier spectra on the whole (radius, latitude) domain (see text). The vertical grey bar shows the uncertainty on the cycle period based on the Fourier spectrum.

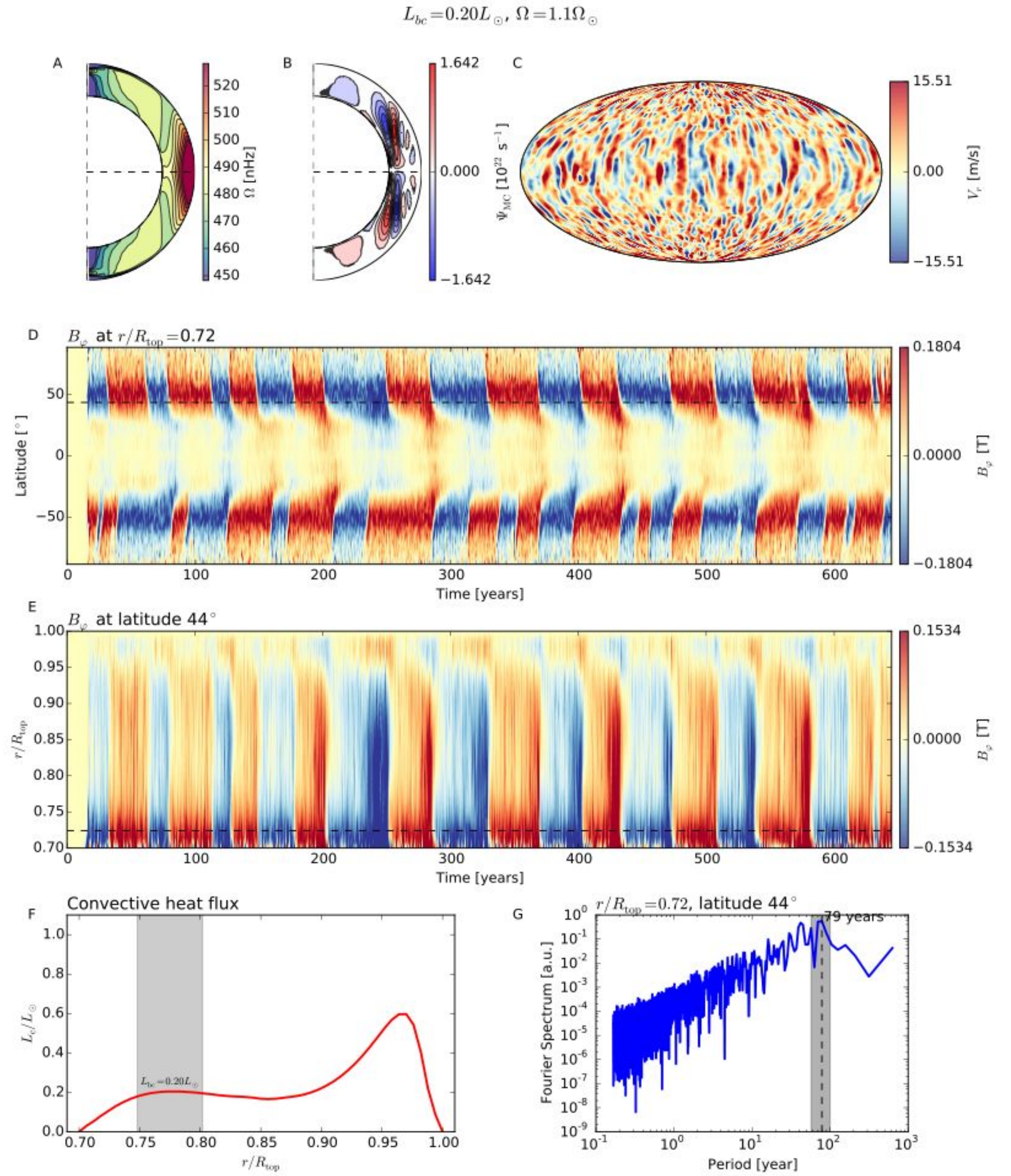

**Fig. S2 Summary of a cyclic numerical simulation.** As in Fig S1, but for case 2.

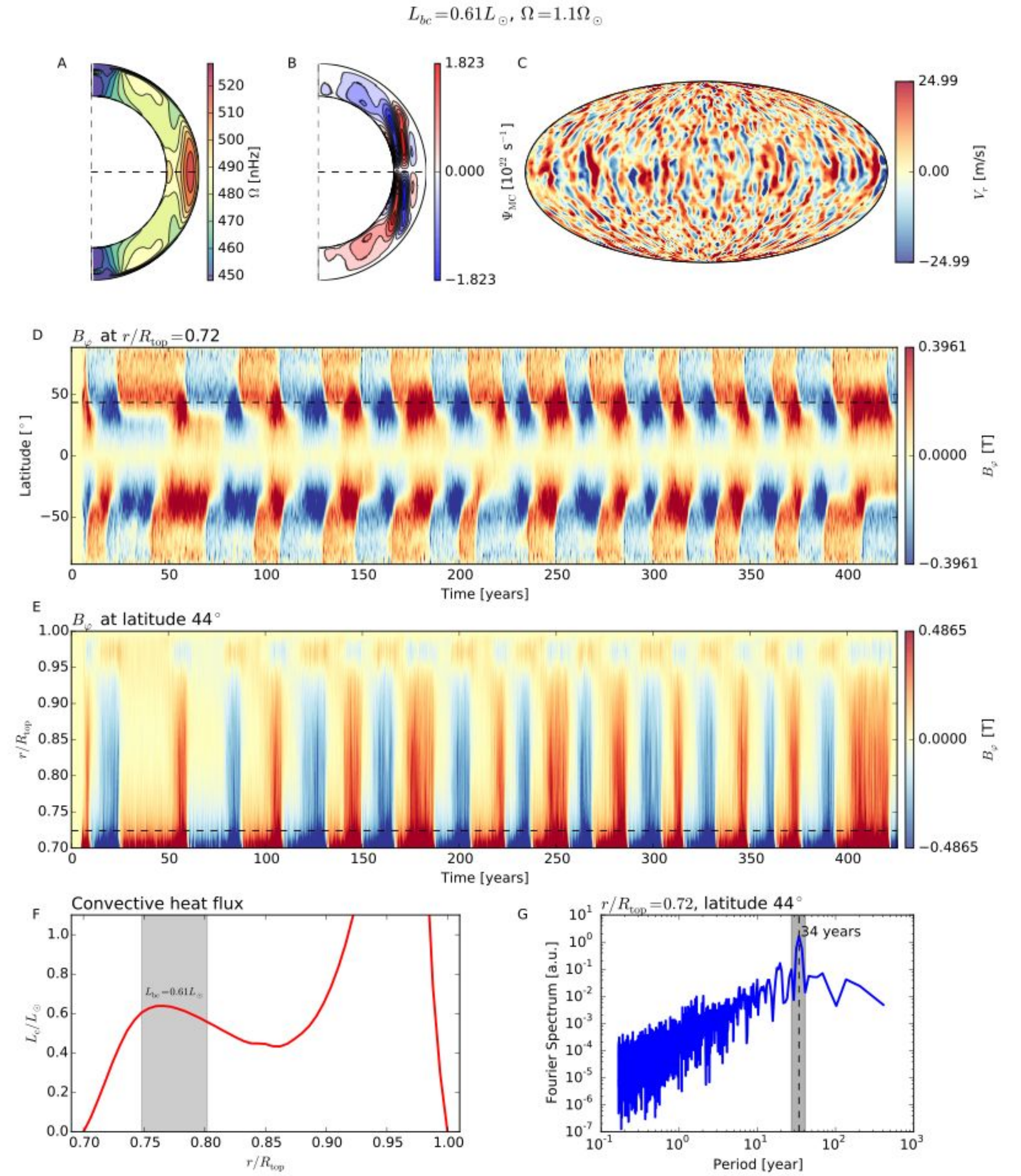

**Fig. S3 Summary of a cyclic numerical simulation.** As in Fig S1, but for case 3.

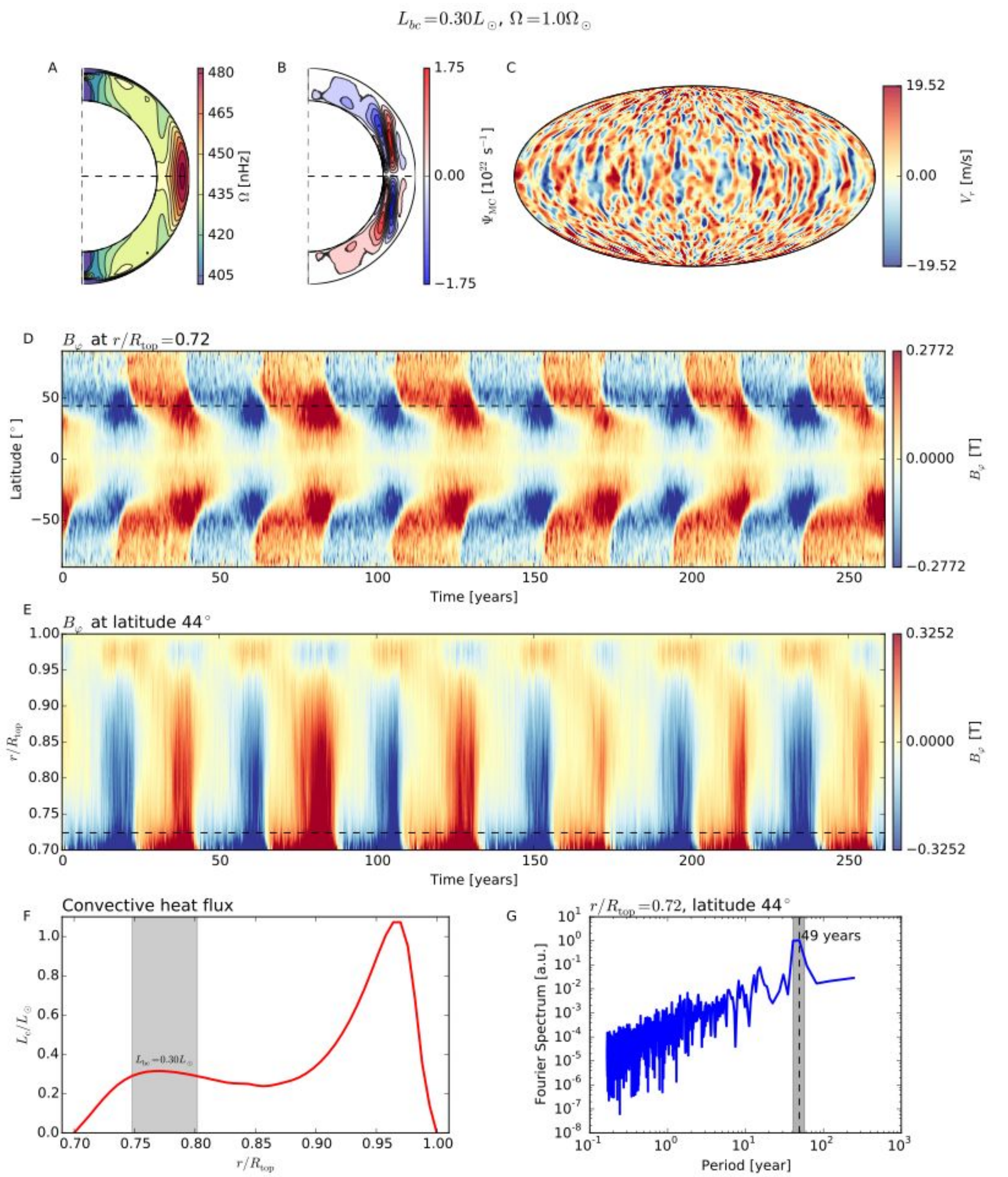

**Fig. S4 Summary of a cyclic numerical simulation.** As in Fig S1, but for case 4.

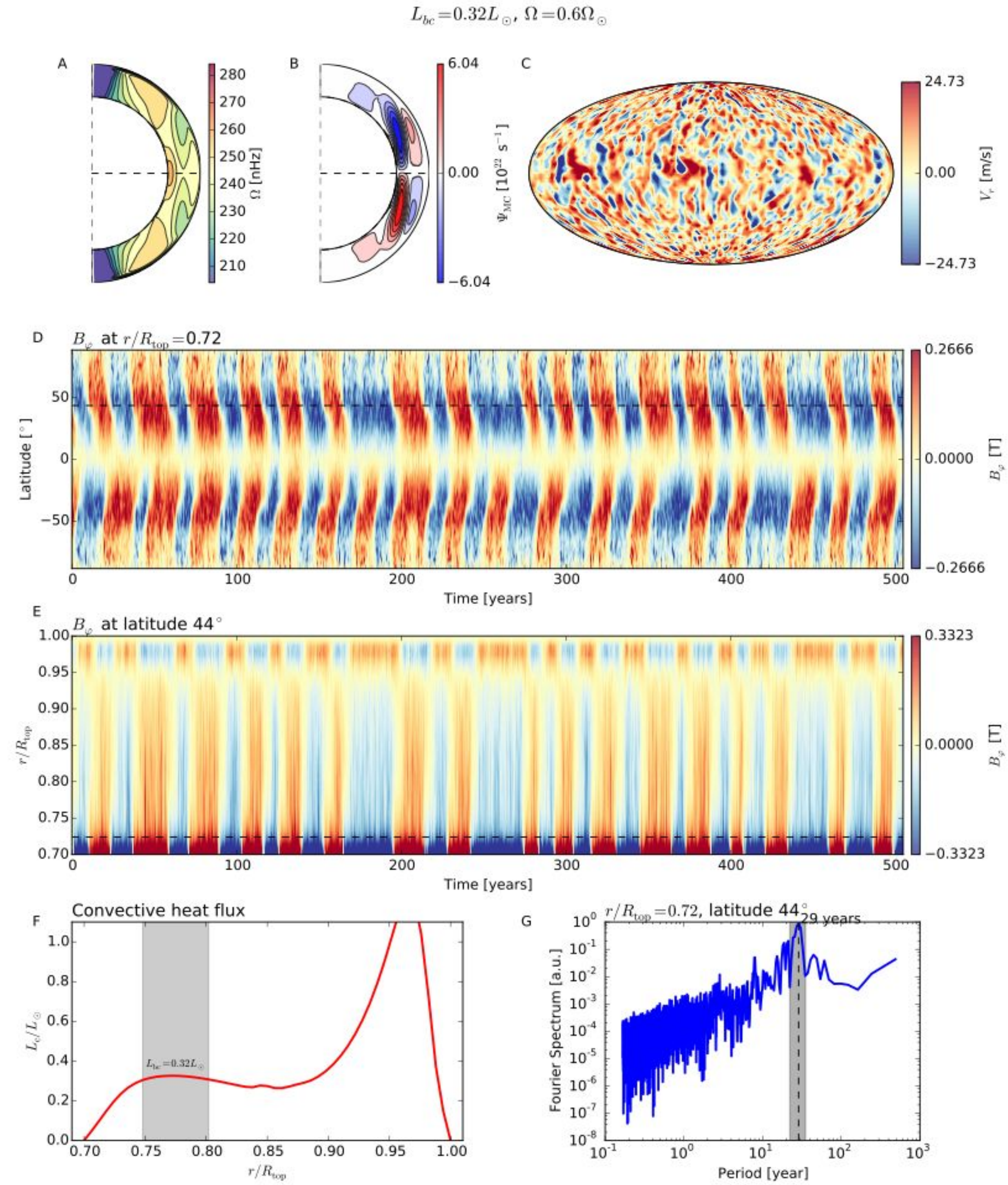

**Fig. S5 Summary of a cyclic numerical simulation.** As in Fig S1, but for case 5.

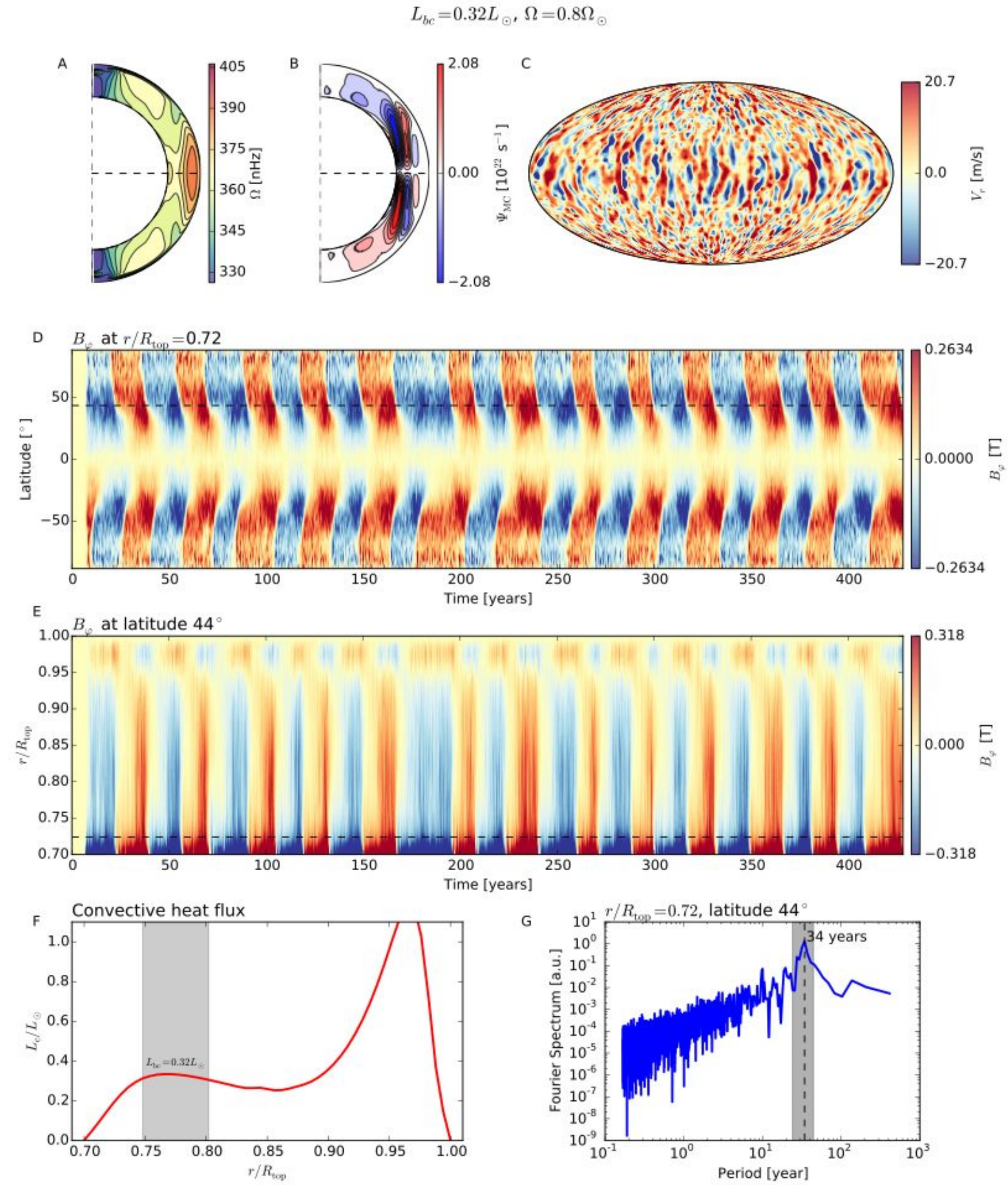

**Fig. S6 Summary of a cyclic numerical simulation.** As in Fig S1, but for case 6.

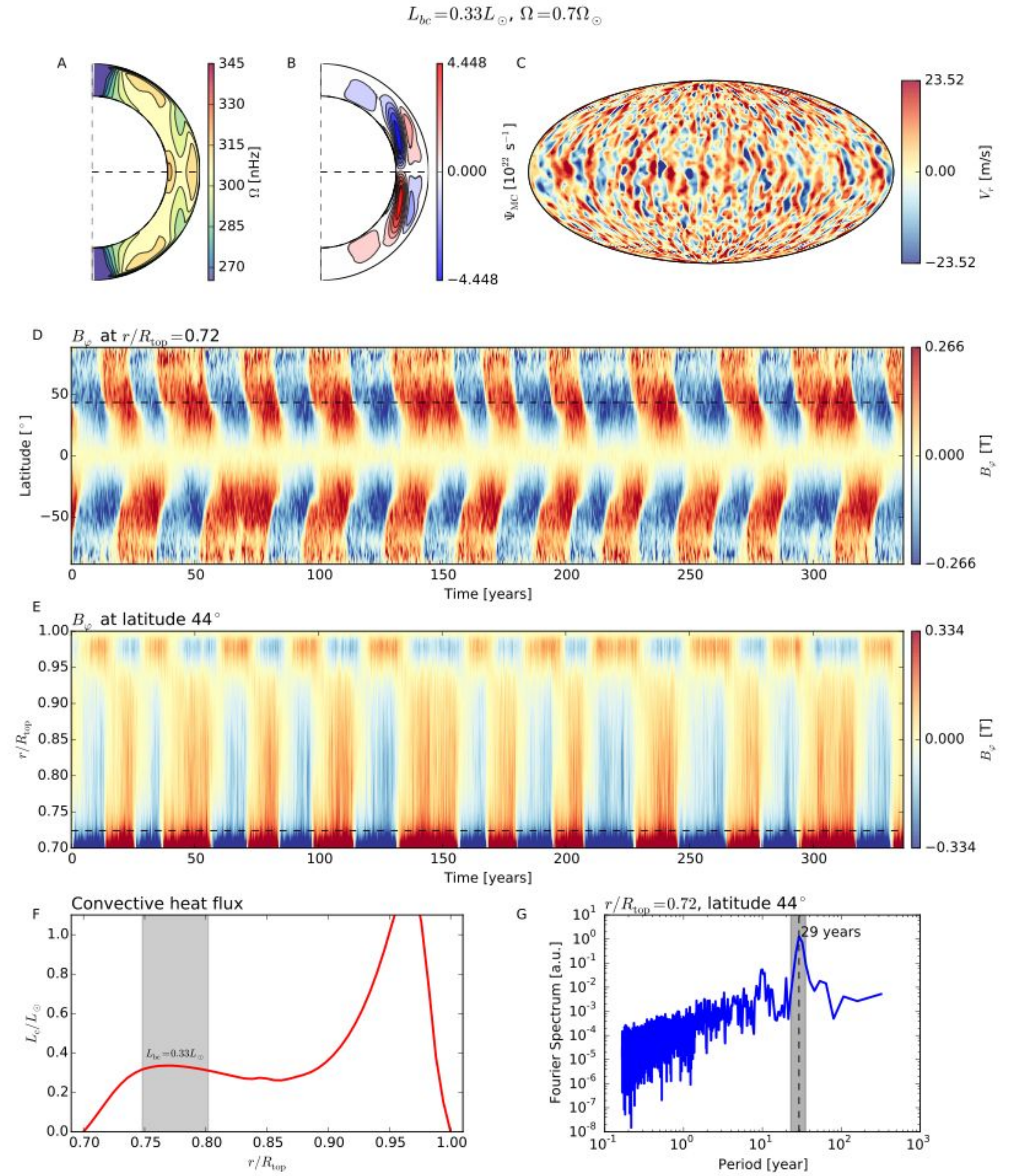

**Fig. S7 Summary of a cyclic numerical simulation.** As in Fig S1, but for case 7.

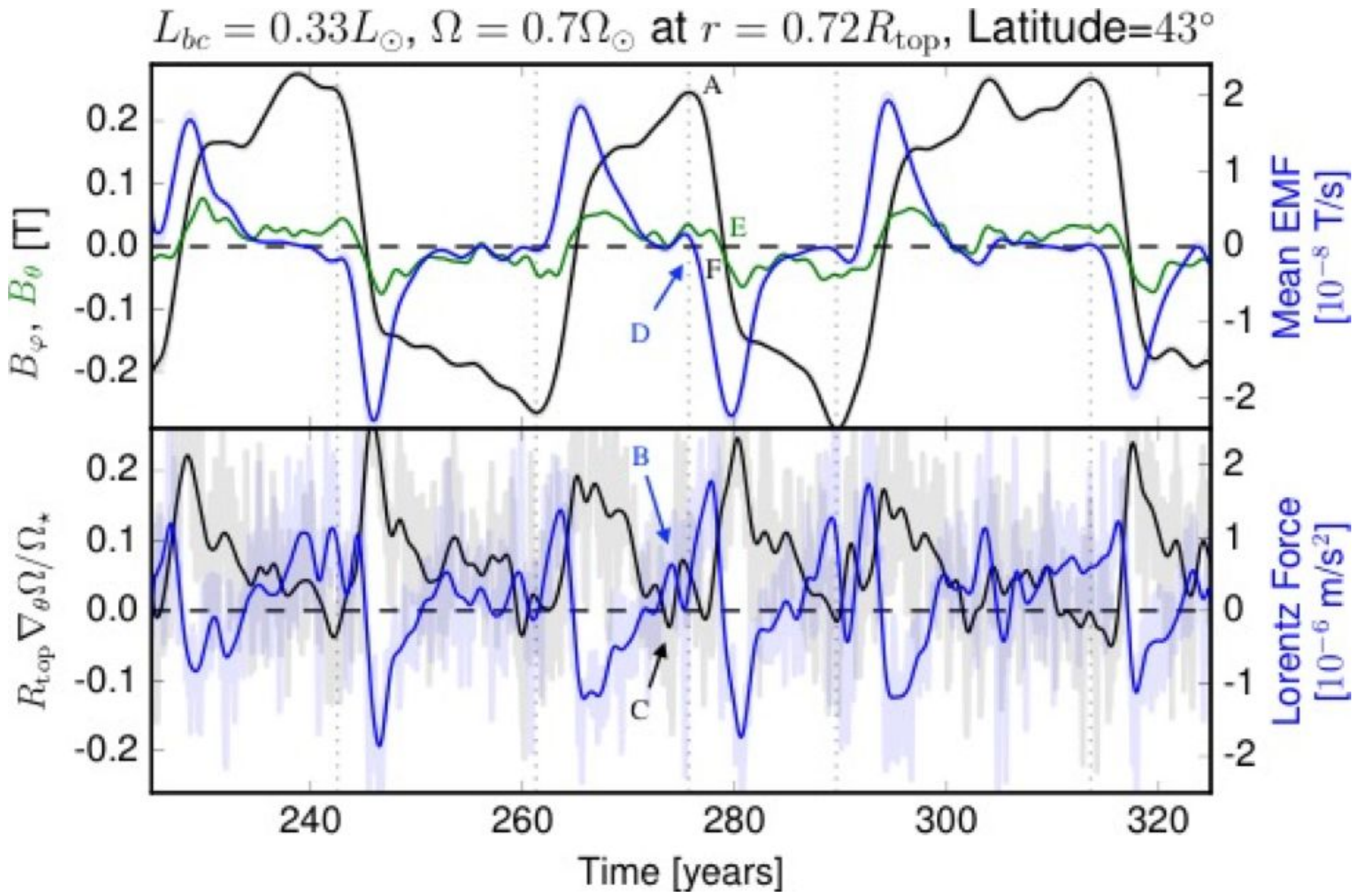

**Fig. S8 Non-linear feedbacks in the magnetic cycles (case 5).** All quantities are averaged over longitude and traced at r=0.72 $R_{top}$ and latitude 43$^\circ$ against time. The upper panel shows the azimuthal (black) and latitudinal (green) components of the magnetic field along with the azimuthal mean electro-motive force (blue). Maxima of the azimuthal magnetic field are labeled by the vertical dashed lines. The lower panel shows the latitudinal gradient of the differential rotation Ω (black) along with the mean Lorentz force (blue) that participates in the evolution of Ω. All fields have been smoothed with a running average of 4 years using a hann window. The un-smoothed fields are shown in the background by the faint thick lines. Letter labels (A to F) describe the non-linear feedback of the large-scale field on the differential rotation (see text).

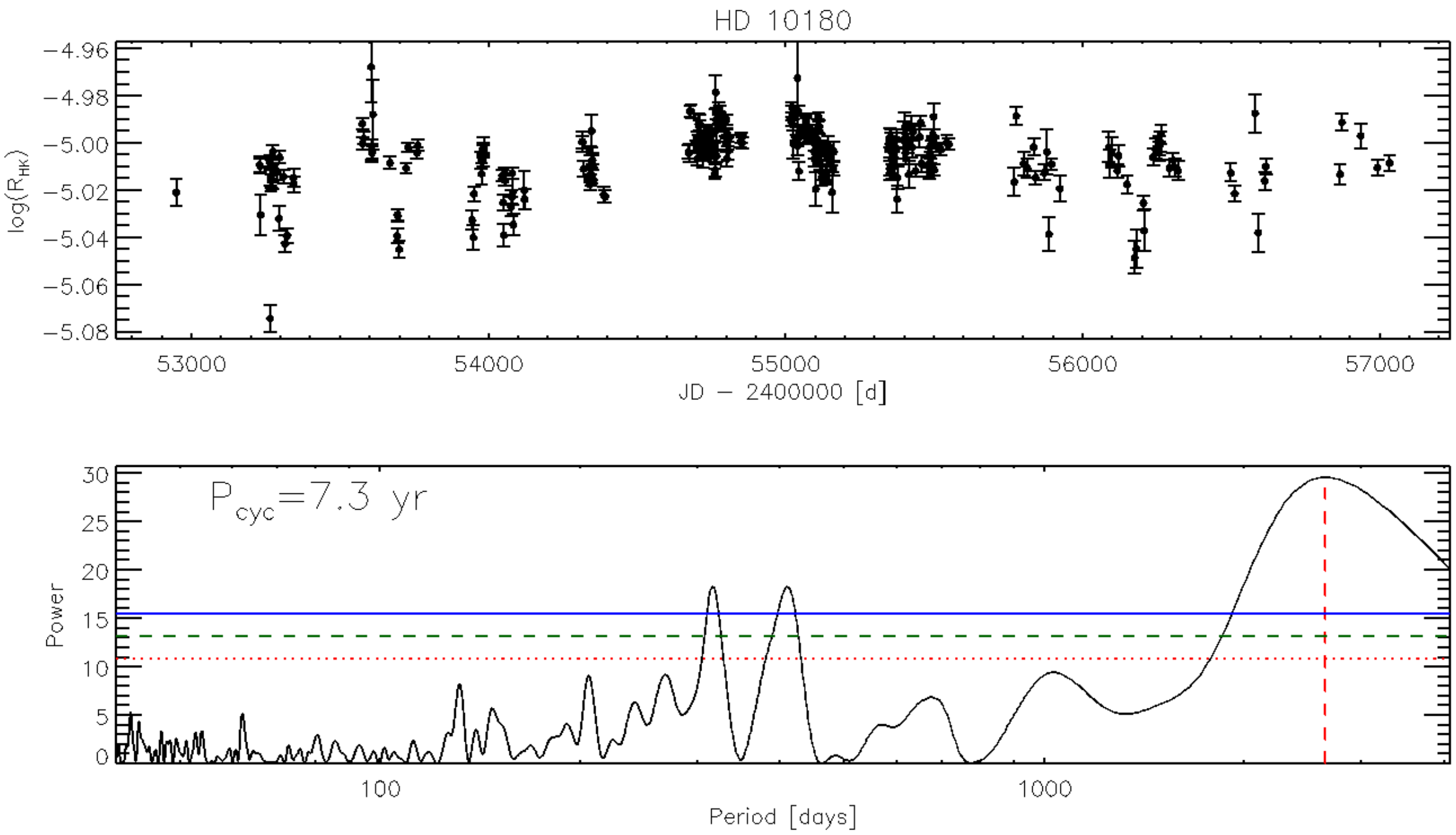

**Fig. S9 Magnetic activity cycle of HD 10180.** Upper panel: activity index Log $R'_{HK}$ defined in (*18*) and computed from the European Southern Observatory (ESO) HARPS southern archive data of the Ca II H & K line cores bands. The abscissa is shown in Julian days. Bottom panel: Lomb-Scargle periodogram(*45*, *46*) of Log $R'_{HK}$ (black line). The peak of the periodogram is identified by the vertical red dashed line and corresponds to a cycle period of 7.3 years. The horizontal red (dotted), blue and green (dashed) lines correspond to the false alarm probability (FAP) of 10%, 1% and 0.1% p-values levels, respectively.